\documentstyle[twocolumn,tighten,aps,epsfig]{revtex}
\begin{document}
\def\seteps#1#2#3#4{\vskip#3\relax\noindent\hskip#1\relax
 \special{eps:#4 x=#2, y=#3}}
\def\centereps#1#2#3{\vskip#2\relax\centerline{\hbox to#1{\special
  {eps:#3 x=#1, y=#2}\hfil}}}
\title{Opposite Thermodynamic Arrows of Time} 
\author{L. S. Schulman}
\address{Physics Department, Clarkson University, Potsdam, New York
13699-5820}
 \date{\today}
 \maketitle
 \begin{abstract}
A model in which two weakly coupled systems maintain opposite running
thermodynamic arrows of time is exhibited. Each experiences its own
retarded electromagnetic interaction and can be seen by the other. The
possibility of opposite-arrow systems at stellar distances is explored
and a relation to dark matter suggested.
\end{abstract}
\medskip
\pacs{05.20.-y, 05.70.Ln, 95.35.+d}

\narrowtext

\bigskip

The possibility of simultaneous opposite running thermodynamic arrows of
time has been raised on several occasions, for didactic purposes
\cite{wiener}, for general  interest \cite{flood} and to confound by
``obvious" counterexample  \cite{flood}. A difficulty in these
considerations is the absence of a  well-defined framework. For example,
one might argue against opposing  arrows as follows. Let the systems be A
and \hbox{B}. An observer in A  will see a succession of small miracles
in B as eggs uncrack, etc. It  would seem that the tiniest interference
by A, the smallest cry of amazement---transmitted to B---would destroy
the monumental coordination needed for B's reversed arrow. That this
argument is flawed is apparent when one realizes that it is phrased from
A's perspective, and takes as natural that the images from B do not
destroy the coordination that B would attribute to \hbox{A}. But whether
the flaw is correctable or whether the conclusion is that {\it both}
arrows would be destroyed, is less clear.

In \cite{correlating,illustration,timebook} a framework for these
questions was proposed. Here I use that framework to show that small
interaction does {\it not\/} destroy the arrows. The question of whether
the systems can communicate will be touched on. Signals are of interest
because of causal paradoxes. One aspect of communication is
electromagnetic radiation and I will extend the Wheeler-Feynman absorber
theory \cite{absorber} to show that each system has its own retarded
interactions, which appear advanced to the other.

Our usual thermodynamic arrow can be phrased as the fact that when
macroscopic (coarse grained) information is given it can be used, by
averaging over the evolution of all microstates consistent with the
macrostate, to estimate the future, but not (in that way) the past. As
argued in \cite{correlating,illustration,timebook}, an unbiased treatment
of thermodynamic arrow questions can be had by giving macroscopic
information at two times (typically, cosmologically remote). It was found
that despite the non-standard conditioning, arrows emerge, consistent
with a thesis correlating the thermodynamic arrow with the expansion of
the universe \cite{goldajp} (or at least with low entropy states at the
remote eras \cite{timebook}).

In \cite{timebook} it was suggested (p.\ 179) that the 2-time formulation
could be used to study opposing arrows, but the inquiry was dismissed as
``science fiction." However, in a time-symmetric universe this
possibility should be considered (in fact this was a complaint in
\cite{flood}, so defense of the arrow-correlation thesis requires this).
Moreover, as proposed below there is also the possibility of physical
relevance in our present cosmological era.

Given systems A and B (for simplicity taken identical) that interact
slightly, conflicting arrows are established through the following
boundary conditions. In each system there is a concept of macrostate,
defined by coarse grains in phase space. At time-0, A and B are
respectively in $\Delta_{Ai}$ and $\Delta_{Bi}$ ($\subset \Gamma \equiv$
phase space energy surface). At time-T they are in $\Delta_{Af}$ and
$\Delta_{Bf}$. The entropy, $S$, of a grain is the logarithm of its
volume. The conflict is imposed by starting A in a small grain, and
putting little or no constraint on its final state. The opposite is done
for B. (``Start" refers to ``$t$," not to a thermodynamic arrow.) Thus:
$S(\Delta_{Ai})= S(\Delta_{Bf})\ll  S(\Delta_{Af}) = S(\Delta_{Bi})$. For
convenience we set $\Delta_{Bi}=\Delta_{Af}=\Gamma$. The relaxation time
for $\Delta_{Ai}$ to spread within $\Gamma$ is denoted $\tau$.

The equation of motion of a particle $\alpha\in$ A is schematically 
$$
\ddot x_\alpha
      =  \sum_{\gamma\in A}F_{\gamma}(x_\alpha) + \sum_{\gamma\in B}
                    F_{\gamma}(x_\alpha)
      =  F^{(A)}+F^{(B)} 
$$
By hypothesis $F^{(B)}$ is small, but not {\it ultra}-small. Thus
$F^{(B)}$ is not {\it so} weak that it would not destroy an entropy
lowering process (such as the time-reverse of a breaking egg) of a
macroscopic system \cite{notultra}. Now if this were a normal physical
problem one would expect the effect of B on A to shorten the relaxation
time: $F^{(B)}$ would be noise on top of the independent motion of
\hbox{A}. But from the B's perspective we might expect extremely rapid
relaxation, because B's interaction destroys A's ability to shrink
entropy (in the direction of B's arrow). Alternatively one might expect
that there simply would be {\it no} solution to the boundary value
problem. If indeed shrinking is instantaneous or solutions do not exist,
then what was wrong with the argument that suggested a small reduction in
$\tau$? Presumably correlations in the ``noise" would allow the small
$F^{(B)}$ to have large coherent effects.

To decide between these alternatives I have done computer simulations
using variations on dynamical systems used to study ergodicity. As will
be seen, the effect of one system on the other is not at all traumatic.
There is simply a moderate shortening of relaxation times.

Each system, A and B, is an ideal gas of particles evolving under the cat
map \cite{arnold}. This is a measure preserving map of the unit square:
$\phi(x,y)=(x+y,x+2y) \;\hbox{mod}\,1$. A single such system has been
used to illustrate conceptual issues and analytic results are available
\cite{illustration,timebook}. We also use the map, $\psi_\alpha(u,v)
\equiv(u+\alpha v,v)\;\hbox{mod}\,1$. Each point $(x_A,y_A)$ in A has a
corresponding one in B, $(x_B,y_B)$. A time step consists of 3 maps: 1)
$\psi_{\alpha/2}$ applied to $(x_A,y_B)$ and $(x_B,y_A)$ separately; 2)
$\phi$ applied to $(x_A,y_A)$ and $(x_B,y_B)$ separately; 3) repeat \#1
\cite{practice}. 

\def\sqmeas{8.5cm}
\def\precap{5pt}

\begin{figure}[t]
\centerline{\epsfig{file=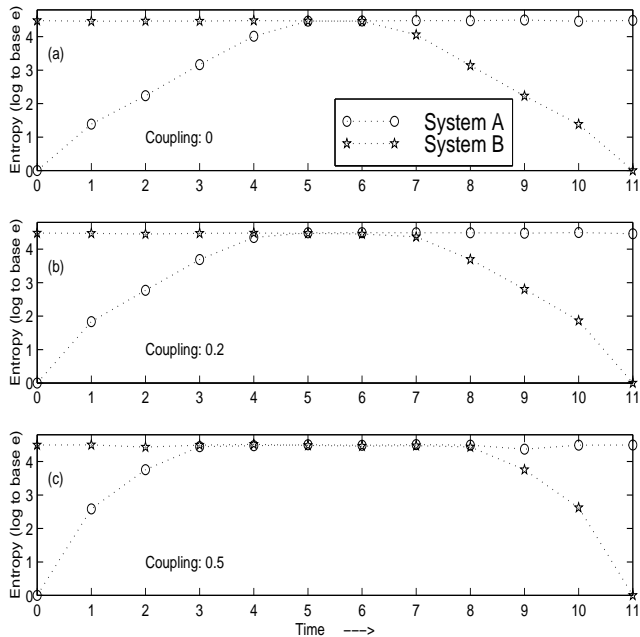,height=\sqmeas,width=\sqmeas}} \vskip \precap
\caption{
Entropy as a function of time for systems, A and B, with opposite
thermodynamic arrows. There are 100 course grains in the unit square and
each simulation uses 500 points. In (a), (b) and (c) the coupling is 0,
0.2, and 0.5 respectively.}
\label{fig1}
\end{figure}

In Fig.\ 1 results are shown for a simulation of 500 pairs of points in
which the initial state of A was confinement in a particular 0.1$\mskip
-2mu\times\mskip1mu$0.1 box with the same final state for \hbox{B}
\cite{method}. Entropies ($S$) of A and B are shown separately, where
$S=-\sum_k\rho_k \log\rho_k$, $k$ labels coarse grains, $N =$ number of
points, and $N\rho_k=$ number of points in grain-$k$. Fig.\ 1a is the
0-coupling result. As expected, the boundary condition gives opposing
arrows. Relaxation times are both about 5. In Fig.\ 1b a coupling
($\alpha$) of 0.2 is used. This conveys the main result of the
simulation, the observation that the two arrows {\it do} persist. What A
feels from B is noise, and the effect is to hasten relaxation. For this
moderate coupling, all that happens is that relaxation takes about 4 time
steps rather than 5. Finally, in Fig.\ 1c $\alpha= 0.5$, for which the
ability of each system to maintain its arrow  is clearly compromised.

We next explore whether the entropy changes yield another property of
arrows, macroscopic causality. By this I mean that effect follows cause,
to be distinguished from microscopic causality, stated, e.g., in terms of
field commutators. Defining a test of (macro) causality requires caution.
Thus with initial-conditions-only the effect of a perturbation is {\it by
definition} subsequent. In \cite{timebook} a consistent test is given by
providing macroscopic data (coarse grains) at two times. The system is
evolved microscopically from initial to final grains with a particular
evolution law, and then again (for the same boundary data) with the same
law on all but one time step, at which time some other law is used. With
low entropy at both ends there are relatively few phase space points
satisfying the boundary conditions. Solution points for perturbed and
unperturbed evolutions are in general different. The test of macroscopic
causality is whether the {\it macroscopic} behavior is different before
the perturbation, after it, or perhaps both \cite{causality}. For our
elaborated cat map, perturbation means that on a particular time step,
instead of applying $\phi$ and $\psi$, another rule is used.

In Fig.\ 2a an entropic history is shown for uncoupled systems. The
perturbation is a faster cat (higher Lyapunov exponent) at time-4
(generated by the matrix [3,2;4,3] (in {\footnotesize MATLAB} notation)).
The entropy, $S(t)$, in the figure is calculated between $t$ and $t+1$.
To better see the effects, in Fig.\ 2b we show only the entropy {\it
change} due to the perturbation. For A the major difference occurs at 4,
while for B it is at 3, consistent with causality. For uncoupled systems
this result is trivial and only shows that our method works. In Fig.\ 2c,
coupling (0.2) is turned on and the same comparison made. Qualitatively
causality persists, although the coupling reduces all deviations.

Understanding radiation with opposing-arrows is no less in need of a
defining framework than our considerations up to now. The language to be
used is time-symmetric electrodynamics and the Wheeler-Feynman absorber
theory \cite{absorber}. Classically there is no loss of generality, since
differences from the standard representation can be eliminated using free
fields. Again consider systems A and B, and write the force on a particle
in, say, A in terms of the advanced and retarded fields of all particles:
$\ddot x_i = \sum_{k\neq i}
\bigl[F_{a}^{(k)}(x_i)+F_{r}^{(k)}(x_i)\bigr]/2$, where $a$ and $r$ refer
to advanced and retarded, respectively, $k\in A\cup B$, and
\hbox{$i\in$~A}. As before, a low entropy macrostate is given for A at
small $t$, high entropy for large $t$, and contrarily for B. As in the
fourth derivation in \cite{absorber}, we rearrange the sum for $\ddot
x_i$, but in a new way:
\begin{eqnarray}
&\ddot x_i =& \sum_{k\in A'} F_{r}^{(k)} 
             +\hbox{$\frac12$}\sum_{k\in A}
              \left[F_{a}^{(k)} - F_{r}^{(k)}\right] \nonumber\\
        &+ &\!\!\sum_{k\in B}  F_{a}^{(k)}
	  +\hbox{$\frac12$}\sum_{k\in B}
                \left[F_{r}^{(k)}-F_{a}^{(k)}\right] \!
        -\hbox{$\frac12$}\left[F_{a}^{(i)}-F_{r}^{(i)}\right]  
\end{eqnarray}
where the prime on $A'$ means $k\neq i$. The term $\bigl[F_{r}^{(i)} -
F_{a}^{(i)}\bigr]/2$ was found by Dirac to give radiation  reaction. We
rewrite Eq.\ (1) in obvious notation
\begin{eqnarray}
\ddot x_i = F_{r}^{(A')} +F_{a}^{(B)}
              + f_{\hbox{\scriptsize rad.\ reac.}}+ E_h
\end{eqnarray}
where $E_h \equiv \frac12 \sum_k
\sigma_k\bigl(F_{a}^{(k)}-F_{r}^{(k)}\bigr)$, $\sigma_k=1$ ($-1$) for
$k\in$ A (B), and is homogeneous (sourceless). These manipulations reduce
to the Wheeler-Feynman calculation when B is empty. They argued that
their $E_h$ was zero, based on the randomness of the particles (this is
the absorber theory). Their explanation of why one should not reverse the
development (to get advanced interactions, etc.) is statistical. In
particular they suppose that the source ($i$) suffers an acceleration.
When only retarded fields are used, they ``had no particular effect on
the acceleration of the source" (\cite{absorber}, p.\ 170). On the other
hand, with a time reversed representation there is coherence in the
source, leading to unlikely behavior. In their words: ``As the result of
chaotic motion going on in the absorber, we see each one of the particles
receiving at the proper moment just the right impulse to generate a
disturbance which converges upon the source at the precise instant when
it is accelerated." As to choosing a representation, they say ``Small
{\it a priori} probability of the given initial conditions provides our
only basis on which to exclude such phenomena."

In our case, for A the unlikely states come at the beginning, for B at
the end. Therefore there should be a different expansion for each. That
is just Eq.\ (2). The key point is that {\it this is still consistent
with electrodynamics}. The field, $E_h$, apparently more complicated
(because of $\sigma_k$) than the one vanishing in \cite{absorber}, is
nevertheless {\it sourceless}.

So it is mathematically consistent for $E_h$ to vanish. Can arguments
like those of Wheeler and Feynman be applied showing that it does in fact
vanish? Since A and B are only weakly coupled this is reasonable. But the
argument could fail if the weak-in-magnitude forces managed peculiar
coherences. It is the point of the numerical simulations reported above
that such correlations do not occur. Those simulations dealt with the
conceptual issues of opposing arrows and although we now have more
complex interactions the conceptual statistical mechanics issues are the
same.

Assuming then that $E_h$ vanishes, what would A see when looking at B?
A's images arise from the advanced field coming from his future.
Successive images present earlier times, as measured by the
causal-entropic arrow of B. Indeed eggs uncrack.

Can this yield causal paradoxes? Can B close the windows and avoid
getting the carpet wet \cite{paradox} because A tells him it's raining
in? In principle such signals could be exchanged and paradoxes avoided as
discussed in \cite{tachyons}. It is also possible that such an
interaction would violate the small coupling assumption. At this stage I
draw no conclusion.

\begin{figure}[t]
\centerline{\epsfig{file=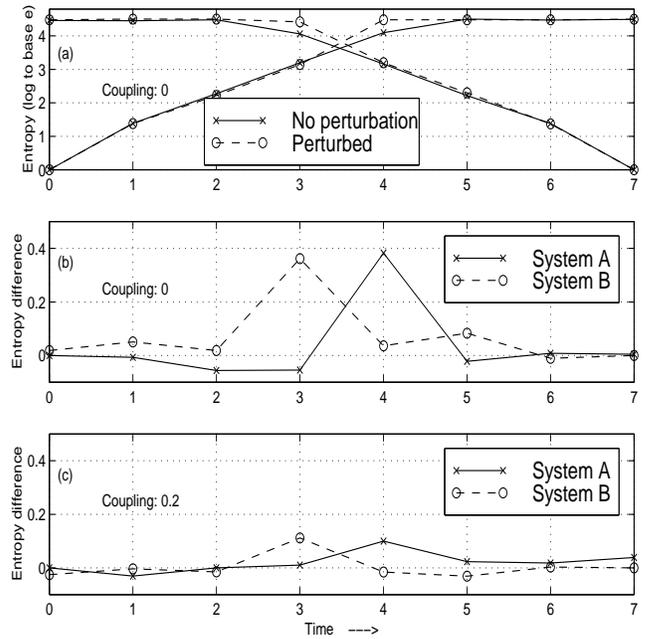,height=\sqmeas,width=\sqmeas}} \vskip \precap
\caption{
Entropy and entropy difference due to perturbation. Coarse grains, etc.\
are as in Fig.\ 1. The solid lines in (a) are (up to statistics) the same
as Fig.\ 1a. For the dashed lines the system is perturbed at $t=4$.
Entropy is calculated {\it between} time steps, so for A ($S\!\uparrow$
for $t\!\uparrow$) the fact that entropy is nearly unchanged for $t\leq4$
means that cause follows effect. For B causality implies that changes
should be at $t<4$, which is confirmed in the figure. For clarity, in (b)
only the difference is shown. Part (c) shows differences for {\it
coupled} systems ($\alpha=0.2$). Again causality is evident, but because
of coupling-induced relaxation the perturbation does not have so marked
an effect.}
\label{fig2}
\end{figure}


Focusing on situations where the small coupling assumption {\it is}
valid, we arrive at the real possibility that at some distance from us
there are regions of opposite running thermodynamic arrows. The extended
absorber theory indicates that we would see them at an era later than our
own, later by the time for light travel to them. How could those regions
have arisen? One possibility is that our universe has a big crunch in the
(our) future and that the other-arrow regions are survivors coming the
other way. If the bang-to-crunch time is long, they would be further away
from their start, hence less likely to have luminous matter. As such, we
would pretty much not see them electromagnetically (but not for the
reasons in \cite{wiener}). On the other hand, there would be no
suppression of gravity. According to this description, this material has
all the properties now attributed to dark matter.

Based on what was learned from the simulations above, there is no bar to
such objects being within our galaxy \cite{temporalcosmo}. Specifically
the radiation from them could be noticeable, but sufficiently weak as not
to overwhelm our normal thermodynamics. A dead star at 50 pc should
satisfy this \cite{halo}. However, this conclusion is not firm, since
with a signal (which gravitational lensing may be) there arises the issue
of whether the small coupling assumption is satisfied (cf.\ the causal
paradoxes). 

Although I have refrained from claiming definite answers to some of the
important questions it is nevertheless clear that at the conceptual level
further progress is possible. In particular, the question of whether
signaling is consistent with weak coupling can be approached by
simulations analogous to but more complicated than what I report above. 

\acknowledgments
I thank A. Ori for urging me to take these matters seriously and for
preliminary discussions, L. J. Schulman for suggestions on dynamical
models, and J. Avron, D. ben-Avraham, P. Facchi, S. Fishman, B. Gaveau,
S. Pascazio, M. Roncadelli and E. Singer for further helpful discussions.
I am grateful to the Inst.\ for Theor.\ Phys.\ and the Petrie Fund of the
Technion and to the NSF (PHY 97 21459) for support.


\begin{references}
\bibitem{notultra} The coupling in the model below has significant effect
on the motion. It is not ``ultra small."
\bibitem{temporalcosmo} This proposal does not depend on the cosmological
scenario given above. ``They" could be there for the same unknown reason
we are here. With a Temporal Cosmological Principle, preferring neither
arrow, there is no reason to rule this out.
\end{references}
\end{document}